\documentclass[12pt]{article}
\usepackage{amsmath, amssymb, epsfig, natbib}
\usepackage{amsmath, bbm, amssymb}
\usepackage{amsthm}
\usepackage{comment}
\usepackage{tabularx}
\usepackage{graphicx, graphics,color}
\usepackage{array}
\usepackage{caption}
\usepackage{booktabs}
\usepackage{float}

\usepackage{listings}
\lstdefinestyle{Rconsole}{
  basicstyle=\ttfamily,
  backgroundcolor=\color{white},
  frame=none,
  showstringspaces=false,
  numbers=none,
  commentstyle=\color{gray}\itshape,
  morecomment=[l]{>},   
  literate={$}{\$}{1}    
}
\usepackage{pdflscape}
\usepackage{enumitem, xcolor}
\usepackage{multirow}
\usepackage{hyperref}
\hypersetup{
	colorlinks=true,        
	citecolor=blue,         
	linkcolor=blue,         
	urlcolor=black,          
	bookmarksnumbered=true,  
	bookmarksopen=true,     
	breaklinks=true,        
	pdfstartview=FitH
}
\usepackage{epstopdf, mathrsfs, framed}

\usepackage{mathtools}
\usepackage{diagbox}
\usepackage[section]{placeins}
\colorlet{linkequation}{blue}
\usepackage{algorithmic}
\usepackage[ruled,linesnumbered,vlined]{algorithm2e}

\newtheorem{remark}{Remark}
\newtheorem{proposition}{Proposition}
\newtheorem{example}{Example}

\usepackage{algorithmic}

\def\prob {{\rm Pr}}
\def\E{\mbox{\rm E}}
\def\var{\mbox{\rm var}}
\def\cov{\mbox{\rm cov}}

\def\O{\mbox{\rm O}}

\def\Orn{\O(n^{-1/2})}

\renewcommand{\vec}[1]{\mbox{\boldmath ${#1}$}}

\def\vmu{\vec{\mu}}

\newcommand{\1}{\mathbf{1}}
\newcommand{\0}{\mathbf{0}}

\date{}

\title{Regional consistency evaluation and sample size calculation under two MRCTs}

\author{Kunhai Qing$^{1}$, Xinru Ren$^{1}$, Shuping Jiang$^{2}$, Ping Yang$^{2}$, \\ Menggang Yu$^{3}$ and Jin Xu$^{1,4,\ast}$}

\begin{document}

\maketitle

\noindent 
$^1$ School of Statistics, East China Normal University, Shanghai, China\\
$^{2}$ BARDS, MSD China, Shanghai, China\\
$^3$ Department of Biostatistics, University of Michigan School of Public Health, USA\\
$^4$ Key Laboratory of Advanced Theory and Application in Statistics and Data Science - MOE, East China Normal University, Shanghai, China\\
$^\ast$ Correspondence to: Jin Xu, Email: {jxu@stat.ecnu.edu.cn}

\begin{abstract}
Multi-regional clinical trial (MRCT) has been common practice for drug development and global registration. The FDA guidance ‘Demonstrating Substantial Evidence of Effectiveness for Human Drug and Biological Products Guidance for Industry’ \citep{FDASEE} requires that substantial evidence of effectiveness of a drug/biologic product to be demonstrated for market approval. In the situations where two pivotal MRCTs are needed to establish effectiveness of a specific indication for a drug or biological product, a systematic approach of consistency evaluation for regional effect is crucial. In this paper, we first present some existing regional consistency evaluations in a unified way that facilitates regional sample size calculation under the simple fixed effects model. Second, we extend the two commonly used consistency assessment criteria of \citet{japan_criteria} in the context of two MRCTs and provide their evaluation and regional sample size calculation. Numerical studies demonstrate the proposed regional sample size attains the desired probability of showing regional consistency. A hypothetical example is presented to illustrate the application. We provide an R package for implementation.  
\end{abstract}


\section{Introduction}\label{sec:intro}
 
In many situations the FDA mandates two adequate and well-controlled trials to establish effectiveness. This reflects the need for substantiation of experimental results, which has often been referred to as the need for replication of the findings \citep{FDASEE}. In the context of global drug development today, these two adequate and well-controlled trials are usually presented as two pivotal multi-region clinical trials (MRCTs) \citep{ICHE17}. MRCTs are conducted under the assumption that the drug demonstrates effectiveness in the overall population and that the same drug effect applies to all regions. Consequently, there are two primary goals of an MRCT: to demonstrate that the drug is effective in the overall population, and to confirm that the drug's effect is consistent across all regions. The second goal is also referred to as the evaluation of regional consistency with the overall drug effect. To date, the overall results of the MRCT have been widely accepted as the primary source of evidence. Meanwhile, evidence of regional consistency is also required by local regulatory agencies (e.g. Pharmaceuticals and Medical Devices Agency of Japan (PMDA) or National Medical Products Administration of China (NMPA)) as a critical component of evidence for a drug to be approved for the local market.
 
In discussion of a regional drug development strategy, a primary consideration is for a region to determine whether and how to join pivotal MRCTs. \cite{ICHE17} encourages participating in MRCTs rather than conducting separate bridging studies to speed up regional drug development. However, it is seldom questioned whether a region should participate in both pivotal MRCTs, as joining only one of the two pivotal MRCTs is still the common practice adopted by many pharmaceutical companies and accepted by local regulatory agencies (e.g., NMPA and PMDA).
Sometimes, participation in both MRCTs may be essential for a region. For example, there are regulatory requirements for a minimum regional subpopulation. The required regional sample size cannot be achieved in one MRCT due to a constrained enrollment window. The strategy for a region to participate in two MRCTs is thus a simple way to mitigate this problem enabling the region to meet the required subpopulation threshold by enrolling additional subjects in the second pivotal MRCT, which may have a longer enrollment window or more enrollment sites. 

This region-join-two-MRCTs strategy raises the question of how to evaluate consistency in a pooled data situation.  
In September 2007, the Japanese Ministry of Health, Labor, and Welfare (MHLW) published the ``Basic Principles on Global Clinical Trials'' guidance related to the planning and implementation of global clinical studies \citep{japan_criteria}. Although the guidance, in a Q\&A format, does not explicitly recommend a specific method for determining the regional sample size in an MRCT for establishing the consistency of treatment effects between the regional population and the overall population, it does provide two methods as examples for recommending the regional sample size. Specifically, the first method assesses the consistency by the probability of the treatment effect of the region of interest exceeding a fraction of the overall effect conditional on the overall effect being significant. The second method assesses the consistency by the probability of all regional treatment effects showing the same trend as the overall effect given the overall significance.  

After the MHLW guidance, the fixed effects model has become the most widely used approach to assess regional consistency. The simple model assumes a uniform underlying treatment effect for all regions and assumes a common variance across regions. Based on the simple model, various criteria for consistency assessment have been proposed and the required regional sample sizes were determined  \citep{ko_criterion,ikeda_sample_2010, tsou_consistency_2011, chen_decision_2012, tsong_assessment_2012, wu2020regional}.   
To introduce greater flexibility for modeling regional effects, the fixed effects model with region-specific treatment effects was adopted, under which variants or extensions of the two criteria of \citet{japan_criteria} and regional sample size calculations were proposed  \citep{chen_assessing_2010, quan_sample_2010}. Moreover, \citet{tanaka_qualitative_2012}, \citet{teng_unified_2017} and \citet{li2024regional} proposed a hypothesis-testing-based approach for consistency assessment.
Alternatively, \citet{quan2013empirical} proposed employing a random effects model and empirical Bayesian shrinkage estimator for the regional effect and then determine the regional sample size accordingly. (See also \citet{adall2021bayes}.)

Yet, there is a lack of methods for assessing the regional consistency in the presence of two MRCTs. A straightforward yet simplistic approach is to apply the consistency evaluation to each of the two MRCTs separately and conclude consistency when consistency is demonstrated in both. However, this approach ignores the trial-wise difference and is inefficient for inference based on the pooled data. In this paper, we first present a unified way for regional consistency evaluation and sample size calculation in one MRCT. Then, we propose new consistency assessment criteria under two MRCTs and derive methods for consistency probability evaluation and sample size determination. These constitute the main results of Section~\ref{sec:method}. Section~\ref{sec:simu} demonstrates the proposed methods achieve the desired operating characteristics through simulation. Section~\ref{sec:app} illustrates the application of the proposed methods to a hypothetical example. Section~\ref{sec:disc} concludes the paper with a discussion of findings and implications.

\section{Method}\label{sec:method}
\subsection{Notation and preliminary}\label{sec:prelim}
Let $Y$ denote the response of the interested endpoint which is either continuous or binary. 
We use superscript $h=`\textrm{t}$' or `c' to indicate the treatment or control group. Let $\mu^{(h)}=\E(Y^{(h)})$ and $\sigma^{2(h)}=\var(Y^{(h)})$.
Let $d=\mu^{(\textrm{t})}-\mu^{(\textrm{c})}$ denote the mean difference. We consider the simple case of independent samples from each group. To test the hypothesis $H_0: d\le 0$ vs $H_1: d>0$, suppose the two-sample $t$-test is adopted with valid assumptions. (It works for both continuous and binary endpoint cases as considered in numerical studies later.) Then, given type-I error $\alpha$ and power $1-\beta$, the required sample sizes (for the overall study) are
\begin{equation}\label{eq:N}
N^{(\textrm{c})}=\frac{\left\{r^{-1}\sigma^{2(\textrm{t})}+\sigma^{2(\textrm{c})}\right\}
(z_{1-\alpha}+z_{1-\beta})^{2}}{d^{2}}, \quad N^{(\textrm{t})}=rN^{(\textrm{c})}
\end{equation}
for the control group and treatment group respectively, where $r$ is the randomization ratio and $z_\alpha$ is the $\alpha$ percentile of the standard normal distribution. Denote $N=N^{(\textrm{t})}+N^{(\textrm{c})}$ as the total sample size.

Suppose there are $K\ (\ge 2)$ regions participating in the trial. For $k=1,\ldots,K$, let $N_{k}^{(h)}$ denote the sample size of group $h$ from region $k$. Clearly, $N^{(h)}=\sum_{k} N_{k}^{(h)}$. Assume the randomization ratio between the treatment group and control group remains the same across regions, i.e., $r=N_{k}^{(\textrm{t})}/N_{k}^{(\textrm{c})}$ for all $k$ (This can be implemented by stratified randomization by region.).  
Let $N_{k}=N_{k}^{(\textrm{t})}+N_{k}^{(\textrm{c})}$ denotes the sample size of region $k$.
Let $f_{k}=N_{k}/N$ be the fraction of samples allocated to region $k$. Our goal is to determine $f_{k}$ for some specific region of interest under a certain consistency criterion.
Here we adopt the fixed effects model and assume the mean difference of the two group is the same across all regions, as in \citet{ikeda_sample_2010} and \citet{chen_decision_2012}.  

For $k=1,\ldots,K$ and $i=1,\ldots,N_{k}^{(h)}$, let $Y_{ki}^{(h)}$ denote the $i$th response from group $h$ in region $k$.
The empirical estimators of $d$ by the pooled samples over all regions and the samples from region $k$ alone are respectively
\begin{equation}\label{eq:D}
D=\frac{\sum_{k,i}Y_{ki}^{(\textrm{t})}}{N^{(\textrm{t})}}-
\frac{\sum_{k,i}Y_{ki}^{(\textrm{c})}}{N^{(\textrm{c})}}, \quad 
D_{k}=\frac{\sum_{i}Y_{ki}^{(\textrm{t})}}{N_{k}^{(\textrm{t})}}-
\frac{\sum_{i}Y_{ki}^{(\textrm{c})}}{N_{k}^{(\textrm{c})}} . 
\end{equation} 
By large sample theory, the distribution of $D$ is approximately
$\mathcal{N}(d,\sigma_d^{2})$, where
\begin{equation}\label{eq:var-D}
\sigma_d^{2}=\var(D)=\frac{(r+1)\left\{\sigma^{2(\textrm{t})}+r\sigma^{2(\textrm{c})}\right\}}{rN}.
\end{equation}
Denote the test statistic for the overall treatment effect by $T=D/\widehat{\sigma}_d$, where $\widehat{\sigma}_d^{2}$ is a consistent estimator of $\sigma_d^{2}$. For example, $\widehat{\sigma}_d^{2}$ is obtained by substituting the sample variances of $\sigma^{2(\textrm{t})}$ and $\sigma^{2(\textrm{c})}$ in (\ref{eq:var-D}). When $T>z_{1-\alpha}$, we reject the null hypothesis and claim the significance of the overall treatment effect.

Moreover, the joint distribution of $(D_{k}, D)^\top$ is approximately normal, 
$\mathcal{N}(\vmu,\Sigma)$, where
\begin{equation}\label{eq:joint-distribution-Dk-D}
\vmu=d(1,1)^\top,\quad
\Sigma=\sigma_d^{2}\begin{pmatrix} f_{k}^{-1} & 1 \\ 1 & 1\end{pmatrix}.
\end{equation}
The covariance of $D_{k}$ and $D$ follows from the fact that $\var(\overline{y})=\cov(\overline{y},\overline{y}_1)$, where $\overline{y}$ and $\overline{y}_1$ are the sample mean of $n$ independent samples of a random variable (with finite variance) and the sample mean of an arbitrary subsample of any size. This fact will be used for later development.  

Finally, denote $\Phi(x)$ and $\phi(x)$ as the distribution function and density function of the standard normal variable. 

\subsection{Regional consistency evaluation and sample size calculation under one MRCT}\label{section:one-study}

To determine $f_{k}$, we first consider the criterion of method I of \citet{japan_criteria}, given by
\begin{equation}\label{eq:criterion-I}
\prob\left(D_{k}\ge \pi D\ |\ T>z_{1-\alpha}\right)>1-\gamma,
\end{equation}
where $\pi\ge 0.5$ represents the fraction of the overall treatment effect is desired to preserve and $1-\gamma\ge 0.8$ is the desired consistency probability (CP). 

We now approximate the consistency probability, i.e., the LHS of (\ref{eq:criterion-I}), in a way that facilitates the regional sample size calculation. It is essentially the same as that given in \citet{ikeda_sample_2010} and \citet{ko_criterion}. 
\begin{proposition}\label{prop:CP-I}
Under the alternative hypothesis with $d>0$, the consistency probability specified by the LHS of (\ref{eq:criterion-I}) is approximately
\begin{equation}\label{eq:criterion-I-expressed}
\frac{1}{1-\beta}\int_{-z_{1-\beta}}^{\infty} \Phi\left(\frac{(1-\pi)(u + z_{1-\alpha}+z_{1-\beta})}{\sqrt{f_{k}^{-1}-1}}\right)\phi(u)du.
\end{equation}
\end{proposition}
\begin{remark}\label{rmk:criterion-I-solve-fk}
It is clear that (\ref{eq:criterion-I-expressed}) increases in $f_{k}$. The numerical integration is straightforward and readily available through function such as \texttt{integrate} of R. Thus, $f_{k}$ can be solved easily by some simple root-finding function such as \texttt{uniroot} of R. Notice that, at the design stage, $f_{k}$ depends only on $\alpha$, $\beta$, $\gamma$ and $\pi$ and is free of $N$ and $K$.  
\end{remark}
\begin{remark}
As showed in the proof in Appendix, the approximation error (between (\ref{eq:criterion-I-expressed}) and LHS of (\ref{eq:criterion-I})) is $\Orn$. The numerical study in Section~\ref{sec:simu} shows the resulting regional sample size from Remark~\ref{rmk:criterion-I-solve-fk} attains the nominal CP up to the third decimal. 
\end{remark}

Second, we consider the criterion of method II of \citet{japan_criteria}, i.e.,
\begin{equation}\label{eq:criterion-II}
\prob\left(D_{k}\geq 0, k=1,\ldots,K\ |\ T>z_{1-\alpha}\right)>1-\gamma,
\end{equation}
which claims the consistency when all regions show the same trend (of directions) after the overall significance. 

Using the same approach as in Proposition~\ref{prop:CP-I}, we get 
\begin{proposition}\label{prop:CP-II}
	Under the alternative hypothesis with $d>0$, the consistency probability specified by the LHS of (\ref{eq:criterion-II}) is approximately
	\begin{equation}\label{eq:criterion-II-expressed}
	\frac{1}{1-\beta}\int_{-z_{1-\beta}}^{\infty} \prod_{k=1}^{K}\Phi\left(\frac{u + z_{1-\alpha}+z_{1-\beta}}{\sqrt{f_{k}^{-1}-1}}\right)\phi(u)du.
	\end{equation}
\end{proposition}
\begin{remark}\label{rmk:criterion-II-solve-fk}
Due to the constraint of $f_1+\cdots+f_{k}=1$, the monotonicity of CP  (\ref{eq:criterion-II-expressed}) with respect to $f_{k}$ becomes complicated.  Nevertheless, CP (\ref{eq:criterion-II-expressed}) is maximized when $f_1=\cdots=f_{k}=K^{-1}$. And the maximum value decreases as $K$ increases. 
\end{remark}
\begin{example}\label{exp:criterion-II}
When $\alpha=0.05$ and $\beta=0.2$, the maximum values of CP are 0.982, 0.897, and 0.772 for $K=2$, 3, and 4, respectively. These upper bounds allow the accommodation of some specified $f_{k}$s of interest in desired ranges. For instance, under $K=3,f_{2}=f_{3}$ and $f_{1}=1-f_{2}-f_{3}$, we only need $f_{1}=10.1\%$ to achieve CP of 80\%. The corresponding empirical CP is 80.1\% (over 10,000 replications). 
\end{example}

When $Y$ is binary, \citet{homma2023cautionary} shows that the approximation (\ref{eq:criterion-II-expressed}) through asymptotic normality loses precision under moderate sample size (since $K$ normality approximations are used collectively). Let $\textrm{Bin}(n,p)$ denote the binomial distribution of parameters $n$ and $p$ and $g(x,n,p)$ denote its associated mass function at $x$. A more accurate evaluation of the consistency probability can be obtained by using an exact method as described in Proposition~\ref{prop:CP-II-bin}. It is noted that for the overall test, the same problem was pointed out by \citet{suissa_85} and \citet{haber_86}. And both recommended using the exact method to calculate the same size.
\begin{proposition}[\citet{homma2023cautionary}]\label{prop:CP-II-bin}
	Under the alternative hypothesis with $d>0$, when $Y$ is binary, the consistency probability specified by the LHS of (\ref{eq:criterion-II}) can be expressed as
	\begin{equation}\label{eq:criterion-II-bin-expressed}
	\frac{1}{1-\beta}\sum_{\left(\sum_{k=1}^{K}u_{k},\sum_{k=1}^{K}v_{k}\right)\in S(z_{1-\alpha})}\prod_{k=1}^{K}
	I(u_{k}>rv_{k})g(u_{k},f_{k}N^{(\textrm{t})},p^{(\textrm{t})}) g(v_{k},f_{k}N^{(\textrm{c})},p^{(\textrm{c})}) ,
	\end{equation}
where $u_{k}\sim \textrm{Bin}(f_{k}N^{(\textrm{t})},p^{(\textrm{t})})$, $v_{k}\sim \textrm{Bin}(f_{k}N^{(\textrm{c})},p^{(\textrm{c})})$, and 
	\begin{equation}\label{eq:criterion-II-bin-expressed-rejection-region}
		S(z_{1-\alpha})=\left\{ \left(u,v\right):\frac{u}{N^{(\textrm{t})}}-\frac{v}{N^{(\textrm{c})}} > z_{1-\alpha}\left[\frac{u\left(N^{(\textrm{t})}-u\right)} {N^{(\textrm{t})^3}}+\frac{v\left(N^{(\textrm{c})}-v\right)} {N^{(\textrm{c})^3}}\right]^{1/2}\right\}.
	\end{equation}
\end{proposition}
\begin{remark}
The direct computation of (\ref{eq:criterion-II-bin-expressed}) through enumeration (over $S(z_{1-\alpha})$) is quite involved. One can instead use Monte Carlo method to compute the LHS of (\ref{eq:criterion-II}). 
\end{remark}

\begin{remark}\label{rmk:criterion-II-bin}
Unlike the normal approximation in Proposition~\ref{prop:CP-II}, the computation by the exact method for the binary response requires specification of the mean value of two groups, i.e.,  $p^{(t)}$ and $p^{(c)}$ (rather than their difference alone), and the randomization ratio.  
\end{remark}

\begin{example}\label{exp:criterion-II-bin}
Fix $\alpha=0.05$, $\beta=0.2$, and $r=1$. Suppose $K=3$, $f_{2}=f_{3}$ and $f_{1}=1-f_{2}-f_{3}$ as in Example~\ref{exp:criterion-II}. For $(p^{(t)},p^{(c)})=(0.8,0.7)$ and $(0.7,0.6)$,  we need $f_{1}=14.9\%$ and 14.0\% to achieve CP of 80\%, respetively. (The corresponding empirical CP is 80.3\% and 80.2\%, respectively.) 
In contrast, the resulting $f_1=10.1\%$ by the normal approximation in Example~\ref{exp:criterion-II} would lead to a deflated CP about 74.7\% and 75.6\%, respectively. 
\end{example}

\subsection{Regional consistency evaluation and sample size calculation under two MRCTs}\label{section:two-studies}

Suppose there are two pivotal, independent MRCTs as described in Section \ref{sec:intro}. For simplicity, consider the scenario where the two MRCTs share the same primary endpoint. We use the superscript `1' and `2' to indicate the study. For $s=1,2$, let $N^{(s)}$ denote the sample size of study $s$. Note that $N^{(1)}$ and $N^{(2)}$ do not have to be the same. For instance, the two studies have an identical primary endpoint, but each is powered by distinct sets of endpoints (e.g., same primary endpoint but different secondary endpoint(s)). Denote the randomization ratio between treatment group and control group as $r^{(s)}=N^{(\textrm{t},s)}/N^{(\textrm{c},s)}$. Like the one study case, assume it is the same across regions.  
We allow the regional fractions to vary across studies. For example, study 1 enrolls more patients from Asia and study 2 enrolls more patients from outside Asia. Denote $f_{k}^{(s)}=N_{k}^{(s)}/N^{(s)}$ for $s=1$ and 2. The other notations extended from the one study case are straightforward and consistent with the one-study case.
Assume the type-I error is fixed the same at $\alpha$ across studies. Yet, the powers can be different as $1-\beta_{1}$ and $1-\beta_{2}$. 

The empirical overall estimator and regional estimators for $d$ from the pooled samples of two studies are respectively given as
\begin{align}\label{eq:diffRates-pool}
D_\textrm{pool} =\frac{\sum_{k,i,s}Y_{ki}^{(\textrm{t},s)}}{\sum_s N^{(\textrm{t},s)}}-
\frac{\sum_{k,i,s}Y_{ki}^{(\textrm{c},s)}}{\sum_s N^{(\textrm{c},s)}}, \quad
D_{k,\textrm{pool}}=\frac{\sum_{i,s}Y_{ki}^{(\textrm{t},s)}}{\sum_s N_{k}^{(\textrm{t},s)}}-
\frac{\sum_{i,s}Y_{ki}^{(\textrm{c},s)}}{\sum_s N_{k}^{(\textrm{c},s)}} .  
\end{align}

We propose to extend criterion (\ref{eq:criterion-I}) to
\begin{equation}\label{eq:criterion-I-pool}
\prob\left(D_{k,\textrm{pool}}\ge \pi D_\textrm{pool}\ |\ T^{(1)}>z_{1-\alpha},T^{(2)}>z_{1-\alpha}\right)>1-\gamma
\end{equation}
to ensure consistency based on the pooled samples from two studies following the establishment of overall significance in both studies. This pooled data-based approach appears more intuitive and reasonable than claiming consistency when the two studies meet the consistency criterion (\ref{eq:criterion-I}) simultaneously. Additional remarks are provided in Remark~\ref{rmk:pooling}.  
\begin{proposition}\label{prop:CP-I-pool} 
	Under the alternative hypothesis with $d^{(1)}>0$ and $d^{(2)}>0$, the consistency probability specified by the LHS of (\ref{eq:criterion-I-pool}) is approximately
\begin{align}\label{eq:criterion-I-pool-expressed}
&\frac{1}{(1-\beta_{1})(1-\beta_{2})}\int_{-z_{1-\beta_{1}}}^{\infty}\int_{-z_{1-\beta_{2}}}^{\infty} \notag\\ 
&\quad
\Phi\left(\frac{(1-\pi)\left\{w^{(1)}\sigma^{(1)}_{d}u+w^{(2)}\sigma^{(2)}_{d}v + w^{(1)}d^{(1)} + w^{(2)}d^{(2)}\right\}}{\sqrt{\left\{\left(f_{k}^{(1)}\right)^{-1}-1\right\} \left(w^{(1)}\sigma^{(1)}_{d}\right)^{2} + \left\{\left(f_{k}^{(2)}\right)^{-1}-1\right\} \left(w^{(2)}\sigma^{(2)}_{d}\right)^{2}}} \right)\phi(u)\phi(v)dudv,
\end{align}
	where $w^{(s)}=N^{(s)}/(N^{(1)}+N^{(2)})$, $s=1,2$. (Note that $\sigma^{(s)}_{d}$ depends on $r^{(s)}$ through (\ref{eq:var-D}).)
\end{proposition}

\begin{remark}\label{rmk:criterion-I-pool} 
	As CP exhibits monotonicity with respect to 
	\[
\zeta=\left(f_{k}^{(1)}\right)^{-1}\left(w^{(1)}\sigma^{(1)}_{d}\right)^{2}+\left(f_{k}^{(2)}\right)^{-1}\left(w^{(2)}\sigma^{(2)}_{d}\right)^{2},
	\]
for given values of $\pi$, $\alpha$, $\beta_{s}$, $d^{(s)}$, $\sigma^{(\mathrm{t},s)}$, $\sigma^{(\mathrm{c},s)}$ and $r^{(s)}$, there are infinitely many pairs of $\left(f_{k}^{(1)},f_{k}^{(2)}\right)$ that can achieve the desired CP. The combined sample size of region $k$, i.e., $ f_{k}^{(1)}N^{(1)} + f_{k}^{(2)}N^{(2)}$, is minimized under the condition 
	\begin{equation}\label{eq:minCond}
		\frac{f_{k}^{(1)}}{f_{k}^{(2)}}= \frac{\sigma^{(1)}_{d}\sqrt{N^{(1)}}}{\sigma^{(2)}_{d}\sqrt{N^{(2)}}}
=\sqrt{\frac{r^{(1)^{-1}}(r^{(1)}+1)\{\sigma^{2(\textrm{t},1)}+r\sigma^{2(\textrm{c},1)}\}}
{r^{(2)^{-1}}(r^{(2)}+1)\{\sigma^{2(\textrm{t},2)}+r\sigma^{2(\textrm{c},2)}\}}}.  
	\end{equation} 
	Note that at the design stage, $f_{k}^{(s)}$ not only depends on $\alpha$, $\beta_{s}$, $\gamma$, but also on $d^{(s)}$, $\sigma^{(\mathrm{t},s)}$, $\sigma^{(\mathrm{c},s)}$ and $r^{(s)}$. 
\end{remark} 
 
\begin{remark}\label{rmk:criterion-I-pool-homo} 
Assuming homogeneous variances, equal treatment effects and equal randomization ratios across two studies, i.e., $\sigma^{2(\textrm{t},1)}=\sigma^{2(\textrm{t},2)}$, $\sigma^{2(\textrm{c},1)}=\sigma^{2(\textrm{c},2)}$, $d^{(1)}=d^{(2)}$ and $r^{(1)}=r^{(2)}$,  (\ref{eq:criterion-I-pool-expressed}) can be simplified as
	\begin{align}\label{eq:criterion-I-pool-expressed-homo} &\frac{1}{(1-\beta_{1})(1-\beta_{2})}\int_{-(z_{1-\beta_{1}}+z_{1-\beta_{2}})/\sqrt{2}}^{\infty}\left[ \Phi\left(\frac{(1-\pi)\left\{\sqrt{2}u + (2z_{1-\alpha}+z_{1-\beta_{1}}+z_{1-\beta_{2}})\right\}}{\sqrt{\left(f_{k}^{(1)}\right)^{-1}+\left(f_{k}^{(2)}\right)^{-1}-2}}\right) \right. \notag \\
		&\qquad\qquad\qquad\qquad \left.\left\{ \Phi\left(u+\sqrt{2}z_{1-\beta_{1}}\right)+\Phi\left(u+\sqrt{2}z_{1-\beta_{2}}\right)-1\right\} \right] \phi(u)du.
	\end{align}   
As pointed out earlier, many pairs of $(f_{k}^{(1)},f_{k}^{(2)})$ can yield a given CP. For example, when $\alpha=0.05$, $\beta_{1} = 0.2$, $\beta_{2} = 0.1$ and $\pi = 0.5$, $\gamma = 0.2$, $(f_{k}^{(1)},f_{k}^{(2)})=(0.141,0.141)$ and $(0.100,0.238)$ can both get CP of 80\%. The combined sample size of region $k$, i.e., $ f_{k}^{(1)}N^{(1)} + f_{k}^{(2)}N^{(2)}$, is minimized under the condition $f_{k}^{(1)} = f_{k}^{(2)}$. 
In this case, $f_{k}^{(s)}$ depends only on $\alpha$, $\beta_{s}$ and $\gamma$. Additionally,  assuming $\beta_{1}=\beta_{2}=\beta$ would simplify (\ref{eq:criterion-I-pool-expressed-homo}) further while the combined sample size of region $k$ is minimized under $f_{k}^{(1)} = f_{k}^{(2)}$.
\end{remark}  
\begin{remark}
Given the CP, additional constraints are required to determine $f_{k}^{(1)}$ and $f_{k}^{(2)}$. For example, when (\ref{eq:minCond}) holds, (\ref{eq:criterion-I-pool-expressed}) increases in $f_{k}^{(2)}$ and $f_{k}^{(2)}$ can be solved easily by a simple root-finding function like \texttt{uniroot} in R. The numerical double integration is available through functions like \texttt{adaptIntegrate} in the R package \texttt{cubature}. 
\end{remark}

\begin{remark}\label{rmk:pooling}
Further rationale is provided for criterion (\ref{eq:criterion-I-pool}). Observe that $D_{k,\textrm{pool}}=w^{(1)}D_{k}^{(1)}+w^{(2)}D_{k}^{(2)}$ and $D_{\textrm{pool}}=w^{(1)}D^{(1)}+w^{(2)}D^{(2)}$. Then,  
\[
\textrm{LHS of}\ (\ref{eq:criterion-I-pool})\ge 
\prob\left(D_{k}^{(1)}\ge \pi D^{(1)}\ |\ T^{(1)}>z_{1-\alpha}\right)\times 
\prob\left(D_{k}^{(2)}\ge \pi D^{(2)}\ |\ T^{(2)}>z_{1-\alpha}\right).
\]
Suppose that $f_{k}^{(1)}=f_{k}^{(2)}=f_{k}$. By the monotonicity of $f_{k}$, criterion (\ref{eq:criterion-I-pool}) based on the pooled samples requires a smaller value for $f_{k}$ than criterion (\ref{eq:criterion-I}) based on two independent samples. For example, given $d^{(s)}=1$, $\sigma^{(\mathrm{t},s)}=\sigma^{(\mathrm{c},s)}=4$, $\beta_{s}=0.2$ and $\alpha=0.05$, to achieve a CP of $\sqrt{80\%}$, ensuring an overall CP of 80\% for the two studies, we need $f_{k}=46.6\%$ for one study with $N=396$. By pooling the samples from the two studies with $N=N^{(1)}+N^{(2)}=396\times 2$, one needs $f_{k}=15.4\%$ to achieve CP 80\% of criterion (\ref{eq:criterion-I-pool}).
\end{remark}

Next, consider the extended criterion of (\ref{eq:criterion-II}) as
\begin{equation}\label{eq:criterion-II-pool}
\prob\left(D_{k,\textrm{pool}}\geq 0, k=1,\cdots,K\ |\ T^{(1)}>z_{1-\alpha},T^{(2)}>z_{1-\alpha}\right)>1-\gamma.
\end{equation}

\begin{proposition}\label{prop:CP-II-pool} 
Under the alternative hypothesis with $d^{(1)}>0$ and $d^{(2)}>0$, the consistency probability specified by the LHS of (\ref{eq:criterion-II-pool}) is approximately
\begin{align}\label{eq:criterion-II-pool-expressed}
	&\frac{1}{(1-\beta_{1})(1-\beta_{2})}\int_{-z_{1-\beta_{1}}}^{\infty}\int_{-z_{1-\beta_{2}}}^{\infty}  \notag\\ 
	&\quad
	\prod_{k=1}^{K}\Phi\left(\frac{ w^{(1)}\sigma^{(1)}_{d}u+w^{(2)}\sigma^{(2)}_{d}v + w^{(1)}d^{(1)} + w^{(2)}d^{(2)} }{\sqrt{\left\{\left(f_{k}^{(1)}\right)^{-1}-1\right\} \left(w^{(1)}\sigma^{(1)}_{d}\right)^{2} + \left\{\left(f_{k}^{(2)}\right)^{-1}-1\right\} \left(w^{(2)}\sigma^{(2)}_{d}\right)^{2}}} \right)\phi(u)\phi(v)dudv.
\end{align} 
\end{proposition}
\begin{remark} When $f_{k}^{(1)}=f_{k}^{(2)}=f_{k}$, the CP in (\ref{eq:criterion-II-pool-expressed}) attains its maximal value at $f_k=K^{-1}$ for all $k$.
\end{remark}
\begin{example}\label{exp:criterion-II-pool} Suppose that $f_{k}^{(1)}=f_{k}^{(2)}=f_{k}$.
For $\alpha=0.05$ and $\beta_{1}=\beta_{2}=0.2$, and assuming homogeneous variances across the two studies as noted in Remark~\ref{rmk:criterion-I-pool-homo}, the maximum values are 0.999, 0.984, and 0.938 for $K=2$, 3, and 4, respectively. When $K=3$, with $f_{2}^{(s)}=f_{3}^{(s)}$ and $f_{1}^{(s)}=1-f_{2}^{(s)}-f_{3}^{(s)}$, one needs only $f_{1}^{(1)}=f_{1}^{(2)}=4.4\%$ to achieve CP of 80\%. The corresponding empirical CP is 80.4\%.  
\end{example}
	
Similar to the one study case, when the response is binary, a more accurate approximation of the CP can be derived as follows. 
\begin{proposition}\label{prop:CP-II-pool-bin}
	Under the alternative hypothesis with $d^{(1)}>0,d^{(2)}>0$, when $Y$ is binary, the consistency probability specified by the LHS of (\ref{eq:criterion-II-pool}) can be expressed as
	\begin{align}\label{eq:criterion-II-pool-bin} 
		\frac{1}{(1-\beta_{1})(1-\beta_{2})}\sum_\Omega &\left\{
 \prod_{k=1}^{K}I\left( \frac{w^{(1)}a^{(1)}_{k}}{f^{(1)}_{k}N^{(\textrm{t},1)}}
 +\frac{w^{(2)}a^{(2)}_{k}}{f^{(2)}_{k}N^{(\textrm{t},2)}}> \frac{w^{(1)}b^{(1)}_{k}}{f^{(1)}_{k}N^{(\textrm{c},1)}}
 +\frac{w^{(2)}b^{(2)}_{k}}{f^{(2)}_{k}N^{(\textrm{c},2)}}\right)\right. \notag \\ 
	& \left.\times \prod_{s} g\left(f^{(s)}_{k}N^{(\textrm{t},s)},a^{(s)}_{k},p^{(\textrm{t},s)}\right)
g\left(f^{(s)}_{k}N^{(\textrm{c},s)},b^{(s)}_{k},p^{(\textrm{c},s)}\right)\right\}
	\end{align}
where 
\[
\Omega=\left\{\left(\sum_{k=1}^{K}a^{(1)}_{k},\sum_{k=1}^{K}b^{(1)}_{k}\right)\in S^{(1)}(z_{1-\alpha})\right\}\bigcap 
\left\{\left(\sum_{k=1}^{K}a^{(2)}_{k},\sum_{k=1}^{K}b^{(2)}_{k}\right)\in S^{(2)}(z_{1-\alpha})\right\},
\]
$a^{(s)}_{k}\sim \textrm{Bin}(f^{(s)}_{k}N^{(\textrm{t},s)},p^{(\textrm{t},s)})$ and $b^{(s)}_{k}\sim \textrm{Bin}(f^{(s)}_{k}N^{(\textrm{c},s)},p^{(\textrm{t},s)})$, and $S^{(s)}(z_{1-\alpha})$ is the same as defined in (\ref{eq:criterion-II-bin-expressed-rejection-region}) for study $s$, $s=1,2$.
\end{proposition}

\begin{example}\label{exp:criterion-II-pool-bin}
Fix $\alpha=0.05$, $\beta_{1}=\beta_{2}=0.2$ and $r^{(s)}=1$.  Suppose that $f_{k}^{(1)}=f_{k}^{(2)}=f_{k}$,  $p^{(\textrm{t},1)}=p^{(\textrm{t},2)}=0.9$, $p^{(\textrm{c},1)}=p^{(\textrm{c},1)}=0.8$. 
Under $K=3$, $f_{2}=f_{3}$ and $f_{1}=1-f_{2}-f_{3}$, we need $f_{1}=6.0\%$ to achieve CP of 80\%.  In contrast, the resulting $f_1$ of 4.4\% by the normal approximation in Example~\ref{exp:criterion-II-pool} would lead to a deflated CP of approximately 77.1\%.
\end{example}

We provide R package for all consistency probability evaluation and regional sample size calculation in GitHub at \href{https://github.com/kunhaiq/ssMRCT.git}{https://github.com/kunhaiq/ssMRCT.git}. We defer the detail to supplementary file.   

\section{Numerical study}\label{sec:simu}
In this section, we conduct simulations to examine the performance of the proposed methods for regional sample size calculation.  

Throughout, we set the nominal consistency probability to be 80\%, i.e. $\gamma = 0.2$, the fraction of overall treatment effect $\pi = 0.5$ and the one-sided $\alpha=0.025$. Consider power $1-\beta$ to take values of  80\% or 90\%. Set the number of replications to be $B=100,000$. We compute the empirical consistency probability (CP) as
\[
\widehat{\textrm{CP}}=\frac{\#\ \textrm{of claims of consistency after rejecting the null hypothesis(es)}}
{\#\ \textrm{of rejections of the null hypothesis(es) out of}\ B\ \textrm{replications}}.
\]
 
\subsection{One study case}

We examine the empirical CP obtained using the calculated regional sample size through (\ref{eq:criterion-I-expressed}) under criterion (\ref{eq:criterion-I}). Set the randomization ratio $r=1$.
(i) For binary response, let the mean of the response under control, denoted by $p^{(\textrm{c})}=\E(Y^{(\textrm{c})})$, take values of 0.5, 0.6, 0.7 or 0.8. And the mean difference $d$ is set to $\{0.1,0.15,0.2\}$ for $p^{(\textrm{c})}=0.5,0.6$ or $0.7$, whereas for $p^{(\textrm{c})}=0.8$, $d$ is constrained to $\{0.1,0.15\}$ since $p^{(\textrm{t})}$ can not be 1. (ii) For continuous response, set $\sigma^{2(\textrm{c})}=\sigma^{2(\textrm{t})}=16$. Let $d$ take values of 1, 1.25 or 2. 
 
Tables~\ref{table:result-one-study-binary} and \ref{table:result-one-study-normal} report the overall sample size obtained from (\ref{eq:N}), the fraction of samples $f_{k}$ calculated from (\ref{eq:criterion-I}) using (\ref{eq:criterion-I-expressed}), and the empirical CP under $N_{k}=f_{k}N$. It is seen that (i) $f_{k}$ is invariant to $N$ under each combination of $\alpha$ and $1-\beta$ as expected and (ii) the calculated fraction of sample can yield the prespecified CP accurately with the average absolute error 0.6\% and 0.5\% over all 22 and 6 rows/cases for the binary response case and continuous response case, respectively. 

The effectiveness of the proposed regional sample size calculation under criterion (\ref{eq:criterion-II}) has already been shown in Example~\ref{exp:criterion-II} for the normal endpoint and in Example~\ref{exp:criterion-II-bin} for the binary endpoint.  

\subsection{Two study case} 

We examine the empirical CP obtained by using the regional sample size through (\ref{eq:criterion-I-pool-expressed}) under criterion (\ref{eq:criterion-I-pool}). (i) For binary response, consider the mean of the response under control, denoted by $p^{(\textrm{c})}=\E\left(Y^{(\textrm{c},1)}\right)=\E\left(Y^{(\textrm{c},1)}\right)$, take values of 0.5 or 0.7. And let the mean difference $d=d^{(1)}=d^{(2)}$ take values of 0.1, 0.15 or 0.2 for both studies. (ii) For continuous response, set $\sigma^{2(\textrm{c},s)}=\sigma^{2(\textrm{t},s)}=16$. Let $d=d^{(1)}=d^{(2)}$ take values of 1, 1.25 or 2 for both studies.  
 
First, set the randomization ratios equal across two studies, i.e., $r^{(1)}=r^{(2)}=1$. Hence the conditions of Remark~\ref{rmk:criterion-I-pool-homo} are satisfied. Tables~\ref{table:result-two-study-homo-binary} and~\ref{table:result-two-study-homo-normal}  report the overall sample sizes obtained from (\ref{eq:N}), the fractions of samples $f_{k}^{(s)}$ solved from (\ref{eq:criterion-I-pool}) using (\ref{eq:criterion-I-pool-expressed-homo}) and the empirical CP with $N_{k}^{(s)}=f_{k}^{(s)}N^{(s)}$, under the constraint that the combined sample size of region $k$ in two studies is minimized, for binary response and normal response, respectively. 
It is seen that (i) $f_{k}$ depends on $p^{(\textrm{c},s)}$ or $d^{(s)}$ (given size and power) as expected and (ii) the calculated fraction of samples can yield the prespecified CP accurately with the average absolute error of 0.7\% and 0.8\% over all 18 and 9 rows/cases for the binary response case and continuous response case, respectively.  

Second, consider different randomization ratios in two studies with $r^{(1)}=1$ and $r^{(2)}=2$.
Tables~\ref{table:result-two-study-different-ratio-binary} and \ref{table:result-two-study-different-ratio-normal} report the overall sample sizes obtained from (\ref{eq:N}), the fractions of samples $f_{k}^{(s)}$ solved from (\ref{eq:criterion-I-pool}) using (\ref{eq:criterion-I-pool-expressed}) and the empirical CP with $N_{k}^{(s)}=f_{k}^{(s)}N^{(s)}$, under the constraint that the combined sample size of region $k$ in two studies is minimized, for binary response and normal response, respectively. 
It is seen that (i) $f_{k}$ depends not only on $p^{(\textrm{c},s)}$ or $d^{(s)}$, but also $r^{(s)}$, under each combination of $\alpha$ and $1-\beta_{s}$ as expected and (ii) the calculated fraction of samples can yield the prespecified CP accurately with the average absolute error of 0.3\% and 0.7\% over all 18 and 9 rows/cases for the binary response case and continuous response case, respectively. 
 
We also conduct simulation for the case of different mean values and variances across studies and find similar results. We defer the detail to supplementary file.   

The effectiveness of the proposed regional sample size calculation under criterion (\ref{eq:criterion-II-pool}) has been demonstrated in Example~\ref{exp:criterion-II-pool} for the normal endpoint and in Example~\ref{exp:criterion-II-pool-bin} for the binary endpoint. 

\section{A hypothetical example}\label{sec:app}

In this section, we provide a hypothetical example, based on real trials, to illustrate the application of the proposed methods. Consider two randomized, double-blind, active-controlled MRCTs designed to evaluate the safety and efficacy of an anti-diabetic agent compared to the standard of care in patients with type 2 diabetes mellitus (HbA1c 7-9.5\%) treated with diet and exercise only  (Study 1) and in patients with type 2 diabetes mellitus (HbA1c 7-10.5\%) inadequately controlled on metformin (Study 2). The second study is conducted in accordance with the guidance document `Demonstrating Substantial Evidence of Effectiveness for Human Drug and Biological Products Guidance for Industry' \citep{FDASEE}. In Study 1, patients are randomized in a 1:1 ratio to be on the once-daily investigational treatment or once-daily active control and then follow up for a 26-week primary treatment period. The primary efficacy objective is the effect of the investigational treatment compared to the active control in lowering HbA1C at Week 26. Study 2 follows the same randomization ratio and primary treatment period as Study 1, followed by an additional 26-week long-term treatment period. The primary efficacy objective for Study 2 includes evaluating the effect of the investigational treatment compared to the active control, both added to background metformin, in lowering HbA1c at Week 26 with HbA1c at Week 52 being one of the secondary objectives. 

The primary efficacy endpoint, common to both studies, is the change from baseline in HbA1c at Week 26 (a continuous endpoint).  This primary endpoint is also the focus of regional evaluation for consistency. In both studies, the assumed treatment effect, the mean difference in primary endpoint values between the treatment and control groups, is 1.2\% with a common standard deviation of 4.0\%. Given a one-sided type I error rate of 0.025 and a power of 90\%, the total sample size needed for each study is 468.

The global enrollment of Study 1 and Study 2 is sequential, with patient recruitment of Study 1 started several months earlier than Study 2. Due to delayed initiation at the regional level, the enrollment window for the region of interest in Study 1 is short, resulting in a limited number of participants recruited from that region. 
Assume the regional authority adopts the criterion (\ref{eq:criterion-I-pool}) for consistency assessment with $\pi=0.5$ based on the pooled data. By the proposed method, if the region of interest contributes an equal proportion of regional participants in both studies (i.e., $f^{(1)}_{k} = f^{(2)}_{k}$), the regional proportions of the overall sample required are 10.9\% and 22.6\% to ensure a CP of 80\% and 90\%, respectively. To more accurately reflect the regional sample imbalances in the two studies, unequal regional proportion pairs such as $(6\%, 60\%)$, $(7\%, 25.7\%)$, $(8\%, 17.4\%)$, and $(9\%, 14.1\%)$ can be planned to achieve a CP of 80\%. Pairs such as $(14\%, 60\%)$, $(16\%, 39.1\%)$, $(18\%, 30.7\%)$, and $(20\%, 26.2\%)$ can be selected to target a CP of 90\%. 

Moreover, if the powers of Studies 1 and 2 are as different as 80\% and 90\%, respectively. Then the overall sample sizes for Studies 1 and 2 become 350 and 468. Under the same assumption of equal proportion of regional participants in both studies, the regional proportions of the overall sample required are 11.8\% and 24.1\% to ensure a CP of 80\% and 90\%, respectively. Similarly, one can use the proposed method to get unequal regional proportion pairs. Pairs of $(7\%, 37\%)$, $(8\%, 22.3\%)$, and $(9\%, 17.0\%)$ can be planned to achieve a CP of 80\%. Pairs such as $(14\%, 87.3\%)$, $(16\%, 49.1\%)$, $(18\%, 36.6\%)$, and $(20\%, 30.4\%)$ can be selected to target a CP of 90\%.

\section{Discussion}\label{sec:disc}
One of the main objectives of an MRCT is to demonstrate that the overall treatment effect applies to all participating regions. To assess such consistency, it is crucial to determine the regional fractions of samples at the design stage according to specific criteria. Although it's common practice for a region of interest to join one of two pivotal MRCTs (a practice accepted by regulatory agencies), there are situations where the region must join both MRCTs.

In this paper, we propose methods for evaluating regional consistency and calculating sample size in two MRCTs based on sensible extensions of the two common criteria established by \citet{japan_criteria}. As the regional fraction usually does not exceed $20\%$ in a single MRCT in practice, the proposed methods offer a practical tool to explore feasible designs. An R package is provided to facilitate application by practitioners.

We also wish to offer two reminders for the use of the proposed design. First, even though we demonstrate the effectiveness of using pooled data to show consistency in Remark~\ref{rmk:pooling}, caution must be taken when there is proven evidence of regional treatment heterogeneity due to covariate distributional shifts \citep{Kulinski:2023} and such shifts differ across studies (e.g., age in study 1 and sex in study 2). In this case, it is not appropriate to use the pooled sample. Second, similar to the relationship between power and sample size, the proposed method calculates the regional fraction depending on the nominal CP. The actual CP may not hold when the sample size falls short. 

The simple fixed effects model with common variance is suitable for many practical applications, as it simplifies method development. However, population heterogeneity across regions due to ethical factors or covariate shifts is also prevalent in reality \citep{Cappellini:2020}. Models that accommodate such variation in regional effects have been proposed \citep{quan2013empirical,quan2014multi}. We will explore novel solutions for the two MRCT problems under those models. Lastly, though we provide a specific method to address the binary response case, there is a need to address challenges related to time-to-event endpoints, especially when the asymptotic normality of the test statistics does not hold. We will report this investigation in a separate work.

\section*{Acknowledgements}
The authors thank Drs. Qiong Shou and William Wang for their comments and suggestions
on the manuscript. The research was supported by a fund of MSD R\&D (China) Co., LTD.

\begin{table}[t]
	\centering
	\caption{Empirical CP obtained under the calculated regional sample size $N_{k}=f_{k}N$ where $N$ is obtained from (\ref{eq:N}) and $f_{k}$ is solved from (\ref{eq:criterion-I}) using (\ref{eq:criterion-I-expressed}) in one study case for binary response }
	\begin{tabular}{*{6}{c}}
        \hline 
        $1-\beta$ & $d$ & $p^{(\textrm{c})}$ & $N$ & $f_{k}$ & CP  \\
        \hline 
        0.8 & 0.1 & 0.5 & 770 & 0.229 & 0.799  \\
          & & 0.6 & 708 & 0.229 & 0.801  \\
          &  & 0.7 & 582 & 0.229 & 0.803  \\
          &  & 0.8 & 394 & 0.229 & 0.808  \\
          & 0.15 & 0.5 & 334 & 0.229 & 0.801  \\
          & & 0.6 & 300 & 0.229 & 0.802  \\
          &  & 0.7 & 236 & 0.229 & 0.805  \\
          &  & 0.8 & 146 & 0.229 & 0.791  \\
          & 0.2 & 0.5 & 182 & 0.229 & 0.810  \\
          & & 0.6 & 158 & 0.229 & 0.809  \\
          &  & 0.7 & 118 & 0.229 & 0.794  \\ 
        \\
        0.9 & 0.1 & 0.5 & 1030 & 0.200 & 0.803 \\
        & & 0.6 & 946 & 0.200 & 0.799  \\
        & & 0.7 & 778 & 0.200 & 0.798 \\
        & & 0.8 & 526 & 0.200 & 0.799 \\
          & 0.15 & 0.5 & 446 & 0.200 & 0.799 \\
        & & 0.6 & 400 & 0.200 & 0.805  \\
        & & 0.7 & 316 & 0.200 & 0.801 \\
        & & 0.8 & 194 & 0.200 & 0.794 \\
          & 0.2 & 0.5 & 242 & 0.200 & 0.808 \\
        & & 0.6 & 212 & 0.200 & 0.808  \\
        & & 0.7 & 158 & 0.200 & 0.793 \\ 
        \hline 
    \end{tabular}
	\label{table:result-one-study-binary}
\end{table}

\begin{table}[t]
	\centering
	\caption{Empirical CP obtained under the calculated the regional sample size $N_{k}=f_{k}N$ where $N$ is obtained from (\ref{eq:N}) and $f_{k}$ is solved from (\ref{eq:criterion-I}) using (\ref{eq:criterion-I-expressed}) in one study case for normal response }
	\begin{tabular}{*{5}{c}}
        \hline 
        $1-\beta$  & $d$ & $N$ & $f_{k}$ & CP  \\
        \hline 
        0.8 &   1 & 504 & 0.229 & 0.802  \\ 
          &  1.5 & 224 & 0.229 & 0.801 \\
          &  2 & 126 & 0.229 & 0.808 \\
        \\ 
        0.9 & 1 & 674 & 0.200 & 0.801 \\ 
          & 1.5 & 300 & 0.200 & 0.804 \\
          & 2 & 170 & 0.200 & 0.808 \\
        \hline 
    \end{tabular}
	\label{table:result-one-study-normal}
\end{table}

\begin{table}[t]
	\centering
	\caption{Empirical CP under the conditions of Remark~\ref{rmk:criterion-I-pool-homo} with the calculated regional sample size $N_{k}^{(s)}=f_{k}^{(s)}N^{(s)}$ where $N^{(s)}$ is obtained from (\ref{eq:N}) and $f_{k}^{(s)}$ is solved from (\ref{eq:criterion-I-pool}) using (\ref{eq:criterion-I-pool-expressed-homo}), subject to minimizing the combined sample size of region $k$ in two studies for binary response }
	\begin{tabular}{*{10}{c}}
        \hline 
        $1-\beta_{1}$ & $1-\beta_{2}$ & $d$ & $p^{(\textrm{c},1)}$ & $p^{(\textrm{c},2)}$ & $N^{(1)}$ & $N^{(2)}$ & $f_{k}^{(1)}$ & $f_{k}^{(2)}$ & CP  \\
        \hline 
        0.8 & 0.8 & 0.1 & 0.5 & 0.5 & 770 & 770 & 0.127 & 0.127 & 0.803  \\ 
        & & & 0.7 & 0.7 & 582 & 582 & 0.127 & 0.127 & 0.804  \\
         & & 0.15 & 0.5 & 0.5 & 334 & 334 & 0.127 & 0.127 & 0.804 \\ 
          & & & 0.7 & 0.7 & 236 & 236 & 0.127 & 0.127 & 0.810\\ 
          & & 0.2 & 0.5 & 0.5 & 182 & 182 & 0.127 & 0.127 & 0.806 \\ 
           & & & 0.7 & 0.7 & 118 & 118 & 0.127 & 0.127 & 0.817 \\ 
        \\
        0.8 & 0.9 & 0.1 & 0.5 & 0.5 & 770 & 1030 & 0.127 & 0.110 & 0.803 \\ 
         & & & 0.7 & 0.7 & 582 & 778 & 0.127 & 0.110 & 0.799 \\
         & & 0.15 & 0.5 & 0.5 & 334 & 446 & 0.127 & 0.110 & 0.807 \\ 
          & & & 0.7 & 0.7 & 236 & 316 & 0.127 & 0.110 & 0.802 \\ 
        & & 0.2 & 0.5 & 0.5 & 182 & 242 & 0.127 & 0.110 & 0.805 \\ 
            & & & 0.7 & 0.7 & 118 & 158 & 0.127 & 0.110 & 0.810 \\ 
        \\
        0.9 & 0.9 & 0.1 & 0.5 & 0.5 & 1030 & 1030 & 0.109 & 0.109 & 0.801 \\ 
        & & & 0.7 & 0.7 & 778 & 778 & 0.109 & 0.109 & 0.800 \\
        & & 0.15 & 0.5 & 0.5 & 446 & 446 & 0.109 & 0.109 & 0.807 \\ 
        & & & 0.7 & 0.7 & 316 & 316 & 0.109 & 0.109 & 0.807 \\ 
        & & 0.2 & 0.5 & 0.5 & 242 & 242 & 0.109 & 0.109 & 0.806 \\ 
         & & & 0.7 & 0.7 & 158 & 158 & 0.109 & 0.109 & 0.814 \\  
        \hline 
    \end{tabular}
	\label{table:result-two-study-homo-binary}
\end{table}

\begin{table}[t]
	\centering
	\caption{Empirical CP under the conditions of Remark~\ref{rmk:criterion-I-pool-homo} with the calculated regional sample size $N_{k}^{(s)}=f_{k}^{(s)}N^{(s)}$ where $N^{(s)}$ is obtained from (\ref{eq:N}) and $f_{k}^{(s)}$ is solved from (\ref{eq:criterion-I-pool}) using (\ref{eq:criterion-I-pool-expressed-homo}), subject to minimizing the combined sample size of region $k$ in two studies for normal response}
		\begin{tabular}{*{8}{c}}
        \hline 
        $1-\beta_{1}$ & $1-\beta_{2}$ & $d$ & $N^{(1)}$ & $N^{(2)}$ & $f_{k}^{(1)}$ & $f_{k}^{(2)}$ & CP  \\
        \hline 
        0.8 & 0.8 & 1 & 504 & 504 & 0.127 & 0.127 & 0.806 \\ 
         & & 1.5 & 224 & 224 & 0.127 & 0.127 & 0.807 \\
         & & 2 & 126 & 126 & 0.127 & 0.127 & 0.817 \\
         \\
        0.8 & 0.9 & 1 & 504 & 674 & 0.127 & 0.110 & 0.800\\ 
         & & 1.5 & 224 & 300 & 0.127 & 0.110 & 0.805 \\
         & & 2 & 126 & 170 & 0.127 & 0.110 & 0.813 \\
         \\
        0.9 & 0.9 & 1 & 674 & 674 & 0.109 & 0.109 & 0.800 \\ 
          & & 1.5 & 300 & 300 & 0.109 & 0.109 & 0.803 \\
          & & 2 & 170 & 170 & 0.109 & 0.109 & 0.806 \\ 
        \hline 
    \end{tabular}
	\label{table:result-two-study-homo-normal}
\end{table}

\begin{table}[t]
	\centering
	\caption{Empirical CP obtained under the calculated regional sample size $N_{k}^{(s)}=f_{k}^{(s)}N^{(s)}$ with where $N^{(s)}$ is obtained from (\ref{eq:N}) and $f_{k}^{(s)}$ is solved from (\ref{eq:criterion-I-pool}) using (\ref{eq:criterion-I-pool-expressed}) with randomization ratios $r^{(1)}=1$ and $r^{(2)}=2$, subject to minimizing the combined sample size of region $k$ in two studies for binary response }
	\begin{tabular}{*{10}{c}}
        \hline  
        $1-\beta_{1}$ & $1-\beta_{2}$ & $d$ & $p^{(\textrm{c},1)}$ & $p^{(\textrm{c},2)}$ & $N^{(1)}$ & $N^{(2)}$ & $f_{k}^{(1)}$ & $f_{k}^{(2)}$ & CP  \\
        \hline  
        0.8 & 0.8& 0.1 & 0.5 & 0.5 & 770 & 873 & 0.123 & 0.131 & 0.804 \\ 
         & & & 0.7 & 0.7 & 582 & 684 & 0.122 & 0.132 & 0.801 \\ 
         & & 0.15 & 0.5 & 0.5 & 334 & 381 & 0.123 & 0.131 & 0.799 \\ 
         & & & 0.7 & 0.7 & 236 & 288 & 0.121 & 0.134 & 0.802 \\ 
         & & 0.2 & 0.5 & 0.5 & 182 & 210 & 0.123 & 0.132 & 0.806 \\ 
         & & & 0.7 & 0.7 & 118 & 153 & 0.120 & 0.135 & 0.803 \\ 
         \\
        0.8 & 0.9 & 0.1 & 0.5 & 0.5 & 770 & 1167 & 0.114 & 0.121 & 0.798 \\ 
         & & & 0.7 & 0.7 & 582 & 915 & 0.113 & 0.122 & 0.800 \\ 
         & & 0.15 & 0.5 & 0.5 & 334 & 510 & 0.114 & 0.121 & 0.804 \\ 
         & & & 0.7 & 0.7 & 236 & 384 & 0.112 & 0.123 & 0.803 \\ 
         & & 0.2 & 0.5 & 0.5 & 182 & 282 & 0.113 & 0.122 & 0.805 \\ 
         & & & 0.7 & 0.7 & 118 & 201 & 0.110 & 0.125 & 0.805 \\ 
         \\ 
        0.9 & 0.9 & 0.1 & 0.5 & 0.5 & 1030 & 1167 & 0.106 & 0.113 & 0.800 \\ 
         & & & 0.7 & 0.7 & 778 & 915 & 0.105 & 0.114 & 0.799 \\ 
         & & 0.15 & 0.5 & 0.5 & 446 & 510 & 0.106 & 0.113 & 0.803 \\ 
         & & & 0.7 & 0.7 & 316 & 384 & 0.104 & 0.115 & 0.800 \\ 
         & & 0.2 & 0.5 & 0.5 & 242 & 282 & 0.105 & 0.113 & 0.804 \\ 
         & & & 0.7 & 0.7 & 158 & 201 & 0.103 & 0.116 & 0.804 \\  
        \hline  
    \end{tabular}
	\label{table:result-two-study-different-ratio-binary}
\end{table}

\begin{table}[t]
	\centering
	\caption{Empirical CP obtained under the calculated regional sample size $N_{k}^{(s)}=f_{k}^{(s)}N^{(s)}$ where $N^{(s)}$ is obtained from (\ref{eq:N}) and $f_{k}^{(s)}$ is solved from (\ref{eq:criterion-I-pool}) using (\ref{eq:criterion-I-pool-expressed}) with randomization ratios $r^{(1)}=1$ and $r^{(2)}=2$, subject to minimizing the combined sample size of region $k$ in two studies for normal response } 
	\begin{tabular}{*{8}{c}}
        \hline 
        $1-\beta_{1}$ & $1-\beta_{2}$ & $d$ & $N^{(1)}$ & $N^{(2)}$ & $f_{k}^{(1)}$ & $f_{k}^{(2)}$ & CP  \\
        \hline 
        0.8 & 0.8  & 1 & 504 & 567 & 0.123 & 0.131 & 0.803 \\  
         & & 1.5 & 224 & 252 & 0.123 & 0.131 & 0.803 \\ 
         & & 2 & 126 & 144 & 0.123 & 0.131 & 0.809 \\ 
         \\
        0.8 & 0.9 & 1 & 504 & 759 & 0.114 & 0.121 & 0.798 \\  
         & & 1.5 & 224 & 339 & 0.114 & 0.121 & 0.803 \\ 
         & & 2 & 126 & 192 & 0.114 & 0.121 & 0.810 \\ 
         \\ 
        0.9 & 0.9 & 1 & 674 & 759 & 0.106 & 0.112 & 0.802 \\  
         & & 1.5 & 300 & 339 & 0.106 & 0.112 & 0.802 \\ 
         & & 2 & 170 & 192 & 0.106 & 0.112 & 0.813 \\  
        \hline 
    \end{tabular}
	\label{table:result-two-study-different-ratio-normal}
\end{table}

\appendix
\section{Proofs}
\subsection{Proof of Proposition~\ref{prop:CP-I}}\label{sec:proof-I}
Let $A=(\begin{smallmatrix} 1 & -\pi \\ 0 & 1\end{smallmatrix})$. The joint distribution of $(D_{k}-\pi D, D)^\top$ is approximately $\mathcal{N}(A\vmu, A\Sigma A^\top)$ with
\[
A\vmu=d(1-\pi,1)^\top,\quad A\Sigma A^\top=\sigma^{2}_d\begin{pmatrix} f^{-1}_{k}-1+(1-\pi)^{2} & 1-\pi \\ 1-\pi & 1 \end{pmatrix}.
\]

The conditional distribution of $D_{k}-\pi D$ given $D$ is approximately $\mathcal{N}(\mu_{r}, \sigma_{r}^{2})$ with
\begin{equation}\label{eq:cond-mean-var}
\mu_{r} = (1-\pi)D,\quad \sigma_{r}^{2} = \sigma^{2}_d (f_{k}^{-1}-1).
\end{equation}
By (\ref{eq:N}) and (\ref{eq:var-D}), we have the identity
\begin{equation}\label{eq:sigma-identity}
	\sigma_d^{2}=\var(D)=\frac{(r+1)\left\{\sigma^{2(\textrm{t})}+r\sigma^{2(\textrm{c})}\right\}}{rN} = \frac{d^{2}}{(z_{1-\alpha}+z_{1-\beta})^{2}}.
\end{equation}

Let $Z=(D-d)/\sigma_{d}$ denote the standardized quantity, which has asymptotically standard normal distribution. Then, we have
\begin{equation}\label{eq:cond-prob-I}
\prob(D_{k}\ge \pi D\ |\ D) \approx \Phi\left(\frac{\mu_{r}}{\sigma_{r}}\right)
=\Phi\left(\frac{(1-\pi)(Z + z_{1-\alpha}+z_{1-\beta})}{\sqrt{f_{k}^{-1}-1}}\right),
\end{equation}
where the second equality follows after substituting (\ref{eq:cond-mean-var}) and (\ref{eq:sigma-identity}).

Express the LHS of (\ref{eq:criterion-I}) as
\begin{equation}\label{eq:condProb-I-ratio}
	\frac{\prob(D_{k}\geq\pi D,T>z_{1-\alpha})}{\prob(T>z_{1-\alpha})}.
\end{equation}
On replacing $\widehat{\textrm{st}}(D)$ by $\textrm{st}(D)$, $\{T>z_{1-\alpha}\}$ is approximately $\{D>z_{1-\alpha}\textrm{st}(D)\}$, which is equivalently to $\{Z>-z_{1-\beta}\}$ by (\ref{eq:sigma-identity}). Then, the numerator of (\ref{eq:condProb-I-ratio}) is approximately the integration of (\ref{eq:cond-prob-I}) over 
$\{Z>-z_{1-\beta}\}$ with respect to $Z$, i.e.,
\[
\int_{Z>-z_{1-\beta}} \Phi\left(\frac{(1-\pi)(Z + z_{1-\alpha}+z_{1-\beta})}{\sqrt{f_{k}^{-1}-1}}\right)d\Phi(Z).
\]
The denominator of (\ref{eq:condProb-I-ratio}) is approximately $\prob(Z>-z_{1-\beta})=1-\beta$.
Substitute these in (\ref{eq:condProb-I-ratio}) to yield (\ref{eq:criterion-I-expressed}).

\subsection{Proof of Proposition~\ref{prop:CP-II}}\label{sec:proof-II}
Extended from the joint distribution of $(D_{k}, D)^\top$ in Section~\ref{sec:prelim}, the joint distribution of $(D_{1},\ldots,D_{K}, D)^\top$ is approximately $\mathcal{N}(\vmu_{2},\Sigma_{2})$ with
\begin{equation}\label{eq:joint-distribution-D1-DK-D}
\vmu_{2}=d\1_{K+1},\quad
\Sigma_{2}=\sigma_d^{2}\begin{pmatrix} f_{1}^{-1} & 0 & \cdots & 1 & 1 \\ 0 & f_{2}^{-1} & \cdots & 0 & 1 \\  \vdots & \vdots & \ddots & \vdots & \vdots \\ 0 & 0 & \cdots & f_{K}^{-1} & 1 \\  1 & 1 & \cdots & 1 & 1\end{pmatrix},
\end{equation}
where $\1_p$ is the $p$-dimension vector of ones. Then, the conditional distribution of $(D_{1},\cdots,D_{K})^{\top}$ given $D$ is approximately  $\mathcal{N}(\vmu_{r}, \Sigma_{r})$ with
\[
\vmu_{r} = D\1_{k},\quad \Sigma_{r} =\sigma_d^{2}\begin{pmatrix} f_{1}^{-1}-1 & 0 & \cdots & 0 \\ 0 & f_{2}^{-1}-1 & \cdots & 0  \\  \vdots & \vdots & \ddots & \vdots  \\ 0 & 0 & \cdots & f_{K}^{-1}-1\end{pmatrix}.
\]

By (\ref{eq:cond-mean-var}), (\ref{eq:sigma-identity}) and $D$ defined in Section~\ref{sec:proof-I}, we have
\begin{equation}\label{eq:cond-prob-II}
	\prob(D_{k}\geq 0, k=1,\cdots,K\ |\ D) \approx \prod_{k=1}^{K}\Phi\left(\frac{D}{\sigma_{d}\sqrt{f_{k}^{-1}-1}}\right) 
= \prod_{k=1}^{K}\Phi\left(\frac{Z + z_{1-\alpha}+z_{1-\beta}}{\sqrt{f_{k}^{-1}-1}}\right).
\end{equation}

Express the LHS of (\ref{eq:criterion-II}) as
\begin{equation}\label{eq:condProb-II-ratio}
	\frac{\prob(D_{k}\geq0, k=1,\cdots,K,T>z_{1-\alpha})}{\prob(T>z_{1-\alpha})}.
\end{equation}
Using the same approximation in the proof of Proposition~\ref{prop:CP-I}, the numerator of (\ref{eq:condProb-II-ratio}) is approximately the integration of (\ref{eq:cond-prob-II}) over $\{Z>-z_{1-\beta}\}$. 
The denominator of (\ref{eq:condProb-II-ratio}) is approximately $\prob(Z>-z_{1-\beta})=1-\beta$.
Substitute these in (\ref{eq:condProb-II-ratio}) to yield (\ref{eq:criterion-II-expressed}).
 
\subsection{Proof of Proposition~\ref{prop:CP-I-pool}}\label{sec:proof-I-pool-fk-diff} 
Let
\[
B=\begin{pmatrix} w^{(1)} & 0 & w^{(2)} & 0 \\ 0 & 1 & 0 & 0 \\
0 & 0 & 0 & 1\end{pmatrix}.
\]
The joint distribution of $\left(D_{k,\textrm{pool}}-\pi D_{\textrm{pool}}, D^{(1)}, D^{(2)}\right)^\top$ is approximately $\mathcal{N}(\vmu_3,\Sigma_3)$ where
\[
\vmu_3=B
\begin{pmatrix}
A\vmu^{(1)} \\
A\vmu^{(2)}
\end{pmatrix},
\quad
\Sigma_3=B\begin{pmatrix}
		A\Sigma^{(1)}A^\top & \0 \\
		\0 & A\Sigma^{(2)}A^\top
		\end{pmatrix}B^\top.
\]
The conditional distribution of $D_{k,\textrm{pool}}-\pi D_\textrm{pool}$ given $D^{(1)}$ and $D^{(2)}$ is $\mathcal{N}(\mu_{r}, \sigma_{r}^{2})$ with
\begin{align*}
	\mu_{r} &= (1-\pi)\left(w^{(1)}D^{(1)}+w^{(2)}D^{(2)}\right), \\
	\sigma_{r}^{2} &= \left\{\left(f_{k}^{(1)}\right)^{-1}-1\right\} \left(w^{(1)}\sigma^{(1)}_{d}\right)^{2} + \left\{\left(f_{k}^{(2)}\right)^{-1}-1\right\} (w^{(2)}\sigma^{(2)}_{d})^{2}.
\end{align*}

For $s=1,2$, by the results from the single study case, 
\begin{equation}\label{eq:sigma-identity-pool} \sigma_d^{2(s)}=\var\left(D^{(s)}\right)=\frac{(r^{(s)}+1)\left\{\sigma^{2(\textrm{t},s)}+r^{(s)}\sigma^{2(\textrm{c},s)}\right\}}{r^{(s)}N^{(s)}} = \frac{d^{2(s)}}{(z_{1-\alpha}+z_{1-\beta^{(s)}})^{2}}.	
\end{equation}
And, let
\[
Z^{(s)}=\frac{D^{(s)}-d^{(s)}}{\sigma^{(s)}_{d}} .
\] 
Similar to the derivation in Section~\ref{sec:proof-I-pool-fk-diff}, we have
\begin{align}\label{eq:cond-prob-I-pool}
&\prob(D_{k,\textrm{pool}}\ge \pi D_\textrm{pool}\ |\ D^{(1)}, D^{(2)}) \notag\\
\approx&  \Phi\left(\frac{\mu_{r}}{\sigma_{r}}\right) = \Phi\left(\frac{(1-\pi)\left(w^{(1)}D^{(1)}+w^{(2)}D^{(2)}\right)}
{\sqrt{\left\{\left(f_{k}^{(1)}\right)^{-1}-1\right\} \left(w^{(1)}\sigma^{(1)}_{d}\right)^{2} + \left\{\left(f_{k}^{(2)}\right)^{-1}-1\right\} \left(w^{(2)}\sigma^{(2)}_{d}\right)^{2}}}\right) \notag\\
	=& \Phi\left(\frac{(1-\pi)\left\{w^{(1)}\sigma^{(1)}_{d}Z^{(1)}+w^{(2)}\sigma^{(2)}_{d}Z^{(2)} + w^{(1)}d^{(1)} + w^{(2)}d^{(2)}\right\}}{\sqrt{\left\{\left(f_{k}^{(1)}\right)^{-1}-1\right\} \left(w^{(1)}\sigma^{(1)}_{d}\right)^{2} + \left\{\left(f_{k}^{(2)}\right)^{-1}-1\right\} \left(w^{(2)}\sigma^{(2)}_{d}\right)^{2}}} \right).
\end{align}

Express the LHS of (\ref{eq:criterion-I-pool}) as
\begin{equation}\label{eq:condProb-I-ratio-pool}
	\frac{\prob\left(D_{k,\textrm{pool}}\geq\pi D_{\textrm{pool}},T^{(1)}>z_{1-\alpha},T^{(2)}>z_{1-\alpha}\right)} {\prob(T^{(1)}>z_{1-\alpha},T^{(2)}>z_{1-\alpha})}.
\end{equation}
On replacing $\widehat{\textrm{st}}(D)$ by $\textrm{st}(D)$, $\left\{T^{(s)}>z_{1-\alpha}\right\}$ is approximately $\left\{D^{(s)}>z_{1-\alpha}\textrm{st}(D^{(s)})\right\}$, which is equivalently to $\left\{Z^{(s)}>-z_{1-\beta_{s}}\right\}$ by (\ref{eq:sigma-identity-pool}). 
Then, the numerator of (\ref{eq:condProb-I-ratio-pool}) is approximately the integration of (\ref{eq:cond-prob-I-pool}) over  
$\left\{Z^{(1)}>-z_{1-\beta_{1}},Z^{(2)}>-z_{1-\beta_{2}}\right\}$. The denominator of (\ref{eq:condProb-I-ratio-pool}) is approximately $\prob\left(Z^{(1)}>-z_{1-\beta_{1}},Z^{(2)}>-z_{1-\beta_{2}}\right)=(1-\beta_{1})(1-\beta_{2})$.
Substitute these in (\ref{eq:condProb-I-ratio-pool}) to get (\ref{eq:criterion-I-pool-expressed}).

Assuming homogeneous variances, equal treatment effects and equal randomization ratios across two studies, i.e., $\sigma^{2(\textrm{t},1)}=\sigma^{2(\textrm{t},2)}$, $\sigma^{2(\textrm{c},1)}=\sigma^{2(\textrm{c},2)}$, $d^{(1)}=d^{(2)}$ and $r^{(1)}=r^{(2)}$. 

Using the transformation $x=(u+v)/\sqrt{2}$ and $y=(u-v)/\sqrt{2}$, express (\ref{eq:criterion-I-pool-expressed}) as  
\begin{align*}
	&\frac{1}{(1-\beta_{1})(1-\beta_{2})}\int_{x>-\frac{1}{\sqrt{2}}(z_{1-\beta_{1}}+z_{1-\beta_{2}})}\int_{x+\sqrt{2}z_{1-\beta_{2}}>y>-x-\sqrt{2}z_{1-\beta_{1}}}  \\
    &\qquad \qquad \qquad \Phi\left(\frac{(1-\pi)\left\{\sqrt{2}x + (2z_{1-\alpha}+z_{1-\beta_{1}}+z_{1-\beta_{2}})\right\}}{\sqrt{\left(f_{k}^{(1)}\right)^{-1}+\left(f_{k}^{(2)}\right)^{-1}-2}}\right)\phi(x)\phi(y)dxdy   \\ 
	=& \frac{1}{(1-\beta_{1})(1-\beta_{2})}\int_{-(z_{1-\beta_{1}}+z_{1-\beta_{2}})/\sqrt{2}}^{\infty}\left[ \Phi\left(\frac{(1-\pi)\left\{\sqrt{2}x + (2z_{1-\alpha}+z_{1-\beta_{1}}+z_{1-\beta_{2}})\right\}}{\sqrt{\left(f_{k}^{(1)}\right)^{-1}+\left(f_{k}^{(2)}\right)^{-1}-2}}\right) \right.  \\
    &\qquad \qquad \qquad \left.\left\{ \Phi\left(x+\sqrt{2}z_{1-\beta_{1}}\right)+\Phi\left(x+\sqrt{2}z_{1-\beta_{2}}\right)-1\right\} \right] \phi(x)dx. 
\end{align*} 
 
\subsection{Proof of Proposition~\ref{prop:CP-II-pool}}\label{sec:proof-II-pool}
By the results of the one study case and independence, we have
\[
(D_{1}^{(1)},\ldots,D_{K}^{(1)},D^{(1)},
D_{1}^{(2)},\ldots, D_{K}^{(2)},D^{(2)})^\top
\stackrel{L}{\longrightarrow} \mathcal{N}\left(
\begin{pmatrix}
\vmu_{2}^{(1)} \\
\vmu_{2}^{(2)}
\end{pmatrix},
\begin{pmatrix}
\Sigma_{2}^{(1)} & \0 \\
\0 & \Sigma_{2}^{(2)}
\end{pmatrix}
\right),
\]
where $\vmu_{2}^{(s)}$ and $\Sigma_{2}^{(s)}$ are defined in (\ref{eq:joint-distribution-D1-DK-D}) for study $s$.
Let
\[
C=\begin{pmatrix} 
	w^{(1)} & 0 & \cdots & 0 & 0 & w^{(2)} & 0 & \cdots & 0 & 0\\
	0 & w^{(1)} & \cdots & 0 & 0 & 0 & w^{(2)} & \cdots & 0 & 0\\
	\vdots & \vdots & \ddots & \vdots & \vdots & \vdots & \vdots & \ddots & \vdots & \vdots\\
	0 & 0 & \cdots &  w^{(1)} & 0 & 0 & 0 & \cdots &  w^{(2)} & 0\\
	0 & 0 & \cdots &  0 & 1 & 0 & 0 & \cdots & 0 & 0\\
	0 & 0 & \cdots &  0 & 0 & 0 & 0 & \cdots & 0 & 1
\end{pmatrix}.
\]
The joint distribution of $(D_{1,\textrm{pool}},\cdots,D_{K,\textrm{pool}}, D^{(1)}, D^{(2)})^\top$ is approximately $\mathcal{N}(\vmu_{4},\Sigma_{4})$ where
\[
\vmu_{4}=C
\begin{pmatrix}
\vmu_{2}^{(1)} \\
\vmu_{2}^{(2)}
\end{pmatrix},
\quad
\Sigma_{4}=C\begin{pmatrix}
		\Sigma_{2}^{(1)} & \0 \\
		\0 & \Sigma_{2}^{(2)}
		\end{pmatrix}C^\top.
\]
The conditional distribution of $(D_{1,\textrm{pool}},\cdots,D_{K,\textrm{pool}})^\top$ given $D^{(1)}$ and $D^{(2)}$ is $\mathcal{N}(\vmu_{r}, \Sigma_{r})$ with
\begin{align*}
	\vmu_{r} &= 
		\left\{w^{(1)}D^{(1)}+w^{(2)}D^{(2)}\right\}\1_{k}, \\ 
	\Sigma_{r} &= \left\{\left(w^{(1)}\sigma^{(1)}_{d}\right)^{2}+\left(w^{(2)}\sigma^{(2)}_{d}\right)^{2}\right\}\begin{pmatrix}
		f_{1}^{-1}-1 & 0 & \cdots & 0 \\ 
		0 & f_{2}^{-1}-1 & \cdots & 0 \\ 
		\vdots & \vdots & \ddots & 0 \\ 
		0 & 0 & \cdots & f_{K}^{-1}-1 
	\end{pmatrix}.
\end{align*}

Similar to the derivation in Section~\ref{sec:proof-II},  we have
\begin{align}\label{eq:cond-prob-II-pool}
&\prob(D_{k,\textrm{pool}}\geq 0, k=1,\cdots,K\ |\ D^{(1)}, D^{(2)}) \notag\\ 
\approx  &  \prod_{k=1}^{K}\Phi\left(\frac{ w^{(1)}D^{(1)}+w^{(2)}D^{(2)} }
{\sqrt{\left\{\left(f_{k}^{(1)}\right)^{-1}-1\right\} \left(w^{(1)}\sigma^{(1)}_{d}\right)^{2} + \left\{\left(f_{k}^{(2)}\right)^{-1}-1\right\} \left(w^{(2)}\sigma^{(2)}_{d}\right)^{2}}}\right)  \notag\\
=&\prod_{k=1}^{K}\Phi\left(\frac{ w^{(1)}\sigma^{(1)}_{d}Z^{(1)}+w^{(2)}\sigma^{(2)}_{d}Z^{(2)} + w^{(1)}d^{(1)} + w^{(2)}d^{(2)} }{\sqrt{\left\{\left(f_{k}^{(1)}\right)^{-1}-1\right\} \left(w^{(1)}\sigma^{(1)}_{d}\right)^{2} + \left\{\left(f_{k}^{(2)}\right)^{-1}-1\right\} \left(w^{(2)}\sigma^{(2)}_{d}\right)^{2}}} \right) .
\end{align}

Express the LHS of (\ref{eq:criterion-II-pool}) as
\begin{equation}\label{eq:condProb-II-ratio-pool}
	\frac{\prob(D_{k,\textrm{pool}}\geq 0,k=1,\cdots,K,T^{(1)}>z_{1-\alpha},T^{(2)}>z_{1-\alpha})} {\prob(T^{(1)}>z_{1-\alpha},T^{(2)}>z_{1-\alpha})}.
\end{equation}
Using the same approximation as in the proof of Proposition~\ref{prop:CP-I-pool}, the numerator of (\ref{eq:condProb-II-ratio-pool}) is approximately the integration of (\ref{eq:cond-prob-II-pool}) over 
$\left\{Z^{(1)}>-z_{1-\beta_{1}},Z^{(2)}>-z_{1-\beta_{2}}\right\}$. 
The denominator of (\ref{eq:condProb-II-ratio-pool}) is approximately $\prob\left(Z^{(1)}>-z_{1-\beta_{1}},Z^{(2)}>-z_{1-\beta_{2}}\right)=(1-\beta_{1})(1-\beta_{2})$.
Substitute these in (\ref{eq:condProb-II-ratio-pool}) to get (\ref{eq:criterion-II-pool-expressed}).

\bibliographystyle{chicago}
\bibliography{TR_ss4twoMRCT}
 
\renewcommand{\thesection}{S\arabic{section}}
\renewcommand{\thefigure}{S\arabic{figure}}
\renewcommand{\thetable}{S\arabic{table}}
\setcounter{section}{0} 
\setcounter{table}{0} 
\setcounter{figure}{0} 

\newpage

\section*{Supplementary Materials for ``Regional consistency evaluation and sample size calculation under two MRCTs''}

\section{Simulation under different mean values and variances across studies }
We present simulation for the cases of different treatment effects and variances of responses across two studies. In particular, for binary response, consider the mean of the response under control as $\E\left(Y^{(\textrm{c},1)}\right)=0.5 $ and $\E\left(Y^{(\textrm{c},2)}\right)=0.7$. Let the mean difference $ d^{(1)}$ and $d^{(2)}$ take values of 0.1 or 0.2.  For continuous response, set $\sigma^{2(\textrm{t},s)}=\sigma^{2(\textrm{c},s)}$ which takes value 16 or 25. Let $d^{(1)}$ and $d^{(2)}$ take values of 1 or 2. 

First, set the randomization ratios equal across two studies, i.e., $r^{(1)}=r^{(2)}=1$. Tables~\ref{table:result-two-study-binary} and~\ref{table:result-two-study-normal}  report the overall sample sizes obtained from (\ref{eq:N}), the fractions of samples $f_{k}^{(s)}$ solved from (\ref{eq:criterion-I-pool}) using (\ref{eq:criterion-I-pool-expressed}) and the empirical CP with $N_{k}^{(s)}=f_{k}^{(s)}N^{(s)}$, under the constraint that the combined sample size of region $k$ in two studies is minimized, for binary response and normal response, respectively. 
It is seen that (i) $f_{k}$ depends on $p^{(\textrm{c},s)}$ or $d^{(s)}$ (given size and power) as expected and (ii) the calculated fraction of samples can yield the prespecified CP accurately with the average absolute error of 0.3\% and 0.3\% over all 4 and 8 rows/cases for the binary response case and continuous response case, respectively.  

Second, consider different randomization ratios in two studies with $r^{(1)}=1$ and $r^{(2)}=2$.
Tables~\ref{table:result-two-study-different-ratio-binary-supp} and \ref{table:result-two-study-different-ratio-normal-supp} report the overall sample sizes obtained from (\ref{eq:N}), the fractions of samples $f_{k}^{(s)}$ solved from (\ref{eq:criterion-I-pool}) using (\ref{eq:criterion-I-pool-expressed}) and the empirical CP with $N_{k}^{(s)}=f_{k}^{(s)}N^{(s)}$, under the constraint that the combined sample size of region $k$ in two studies is minimized, for binary response and normal response, respectively. 
It is seen that (i) $f_{k}$ depends not only on $p^{(\textrm{c},s)}$ or $d^{(s)}$, but also $r^{(s)}$, $\sigma^{2(\textrm{t},s)}$ and $\sigma^{2(\textrm{c},s)}$, under each combination of $\alpha$ and $1-\beta_{s}$ as expected and (ii) the calculated fraction of samples can yield the prespecified CP accurately with the average absolute error of 0.2\% and 0.4\% over all 4 and 8 rows/cases for the binary response case and continuous response case, respectively.
  
\begin{table}[t]
	\centering
	\caption{Empirical CP obtained under the calculated regional sample size $N_{k}^{(s)}=f_{k}^{(s)}N^{(s)}$ where $N^{(s)}$ is obtained from (\ref{eq:N}) and $f_{k}^{(s)}$ is solved from (\ref{eq:criterion-I-pool}) using (\ref{eq:criterion-I-pool-expressed}), subject to minimizing the combined sample size of region $k$ in two studies for binary response }
	\begin{tabular}{*{9}{c}}
        \hline 
        $1-\beta_{1}$ & $1-\beta_{2}$ & $d^{(1)}$ & $d^{(2)}$  & $N^{(1)}$ & $N^{(2)}$ & $f_{k}^{(1)}$ & $f_{k}^{(2)}$ & CP  \\
        \hline 
        0.8 & 0.8 & 0.1 & 0.2 & 770 & 118 & 0.162 & 0.127 & 0.803  \\  
        0.8 & 0.9 & 0.1 & 0.2 & 770 & 158 & 0.151 & 0.118 & 0.800  \\  
         & & 0.2 & 0.1 & 182 & 778 & 0.136 & 0.122 & 0.804  \\  
        0.9 & 0.9 & 0.1 & 0.2 & 1030 & 158 & 0.140 & 0.109 & 0.803  \\  
        \hline 
    \end{tabular}
	\label{table:result-two-study-binary}
\end{table}

\begin{table}[t]
	\centering
	\caption{Empirical CP obtained under the calculated regional sample size $N_{k}^{(s)}=f_{k}^{(s)}N^{(s)}$ where $N^{(s)}$ is obtained from (\ref{eq:N}) and $f_{k}^{(s)}$ is solved from (\ref{eq:criterion-I-pool}) using (\ref{eq:criterion-I-pool-expressed}), subject to minimizing the combined sample size of region $k$ in two studies for normal response}
		\begin{tabular}{*{11}{c}}
        \hline 
        $1-\beta_{1}$ & $1-\beta_{2}$ & $d^{(1)}$ & $d^{(2)}$ & $\sigma^{2(\textrm{c},1)}$ & $\sigma^{2(\textrm{c},2)}$ & $N^{(1)}$ & $N^{(2)}$ & $f_{k}^{(1)}$ & $f_{k}^{(2)}$ & CP  \\
        \hline 
        0.8 & 0.8 & 1 & 2 & 25 & 16 & 786 & 126 & 0.160 & 0.128 & 0.804 \\   
         & & 2 & 1 & 25 & 16 & 198 & 504 & 0.149 & 0.119 & 0.805 \\   
        0.8 & 0.9 & 1 & 2 & 25 & 16 & 786 & 170 & 0.149 & 0.119 & 0.802 \\   
         & & & & 16 & 25 & 504 & 264 & 0.108 & 0.135 & 0.805 \\  
         & & 2 & 1 & 25 & 16 & 198 & 674 & 0.141 & 0.112 & 0.802 \\   
         & & & & 16 & 25 & 126 & 1052 & 0.118 & 0.147 & 0.800 \\  
        0.9 & 0.9 & 1 & 2 & 25 & 16 & 1052 & 170 & 0.138 & 0.110 & 0.801 \\   
         & & 2 & 1 & 25 & 16 & 264 & 674 & 0.128 & 0.102 & 0.803 \\   
        \hline 
    \end{tabular}
	\label{table:result-two-study-normal}
\end{table}

\begin{table}[t]
	\centering
	\caption{Empirical CP obtained under different randomization ratios ($r^{(1)}=1$ and $r^{(2)}=2$) with the calculated regional sample size $N_{k}^{(s)}=f_{k}^{(s)}N^{(s)}$ where $N^{(s)}$ is obtained from (\ref{eq:N}) and $f_{k}^{(s)}$ is solved from (\ref{eq:criterion-I-pool}) using (\ref{eq:criterion-I-pool-expressed}), subject to minimizing the combined sample size of region $k$ in two studies for binary response }
	\begin{tabular}{*{9}{c}}
        \hline 
        $1-\beta_{1}$ & $1-\beta_{2}$ & $d^{(1)}$ & $d^{(2)}$  & $N^{(1)}$ & $N^{(2)}$ & $f_{k}^{(1)}$ & $f_{k}^{(2)}$ & CP  \\
        \hline 
        0.8 & 0.8 & 0.1 & 0.2 & 770 & 153 & 0.151 & 0.134 & 0.800  \\  
        0.8 & 0.9 & 0.1 & 0.2 & 770 & 201 & 0.139 & 0.123 & 0.804  \\  
         & & 0.2 & 0.1 & 182 & 915 & 0.132 & 0.128 & 0.802  \\  
        0.9 & 0.9 & 0.1 & 0.2 & 1030 & 201 & 0.130 & 0.115 & 0.801  \\  
        \hline 
    \end{tabular}
	\label{table:result-two-study-different-ratio-binary-supp}
\end{table}

\begin{table}[t]
	\centering
	\caption{Empirical CP obtained under different randomization ratios ($r^{(1)}=1$ and $r^{(2)}=2$) with the calculated regional sample size $N_{k}^{(s)}=f_{k}^{(s)}N^{(s)}$ where $N^{(s)}$ is obtained from (\ref{eq:N}) and $f_{k}^{(s)}$ is solved from (\ref{eq:criterion-I-pool}) using (\ref{eq:criterion-I-pool-expressed}), subject to minimizing the combined sample size of region $k$ in two studies for normal response } 
	\begin{tabular}{*{11}{c}}
        \hline 
        $1-\beta_{1}$ & $1-\beta_{2}$ & $d^{(1)}$ & $d^{(2)}$ & $\sigma^{2(\textrm{c},1)}$ & $\sigma^{2(\textrm{c},2)}$ & $N^{(1)}$ & $N^{(2)}$ & $f_{k}^{(1)}$ & $f_{k}^{(2)}$ & CP  \\
        \hline 
        0.8 & 0.8 & 1 & 2 & 25 & 16 & 786 & 144 & 0.155 & 0.131 & 0.804 \\   
         & & 2 & 1 & 25 & 16 & 198 & 567 & 0.147 & 0.124 & 0.806 \\   
        0.8 & 0.9 & 1 & 2 & 25 & 16 & 786 & 192 & 0.143 & 0.122 & 0.802 \\   
         & & & & 16 & 25 & 504 & 297 & 0.103 & 0.136 & 0.804 \\  
         & & 2 & 1 & 25 & 16 & 198 & 759 & 0.138 & 0.117 & 0.804 \\   
         & & & & 16 & 25 & 126 & 1185 & 0.114 & 0.151 & 0.801 \\  
        0.9 & 0.9 & 1 & 2 & 25 & 16 & 1052 & 192 & 0.133 & 0.113 & 0.802 \\   
         & & 2 & 1 & 25 & 16 & 264 & 759 & 0.126 & 0.107 & 0.803 \\   
        \hline 
    \end{tabular}
	\label{table:result-two-study-different-ratio-normal-supp}
\end{table}

\section{R package}
We present the use of the main functions of the \textbf{ssMRCT} package. The primary functions for \textbf{ssMRCT} users are \texttt{conProb2} and \texttt{regFrac2}. Specifically, \texttt{conProb2} returns the consistency probability for two MRCTs, and \texttt{regFrac2} returns the regional fraction for two MRCTs.
The code to obtain the regional fractions in Remark~\ref{rmk:criterion-I-pool-homo} is:
\begin{lstlisting}[style=Rconsole]
regFrac2(alpha=0.05,power1=0.8,power2=0.9,d1=1,sigmaTrt1=4)
\end{lstlisting}
The output is 
\begin{lstlisting}[style=Rconsole]
$rF1
[1] 0.1407622

$rF2
[1] 0.1407622

$N1
[1] 396

$N2
[1] 550
\end{lstlisting}

Using the optimal regional fractions $(0.1407622,0.1407622)$ as shown above, we can use \texttt{conProb2} to check the consistency probability: 
\begin{lstlisting}[style=Rconsole]
conProb2(alpha=0.05,power1=0.8,power2=0.9,rF1=0.1407622)
\end{lstlisting} 
The output is 
\begin{lstlisting}[style=Rconsole]
$CP
[1] 0.8000277

$N1
[1] 396

$N2
[1] 550
\end{lstlisting}

\begin{quote}
Detailed description of the arguments \texttt{regFrac2} are as follows:
\begin{description}
	\item \texttt{alpha}: The Type I error.
	\item \texttt{power1}: Power for MRCT 1.
	\item \texttt{power2}: Power for MRCT 2. Defaults to \texttt{power1}.
	\item \texttt{pi}: The threshold ratio in the extended Japan's criterion I (conditional version). Defaults to 0.5.
	\item \texttt{CP}: The consistency probability. Defaults to 80\%. 
	\item \texttt{d1}: The true mean of difference of response for MRCT 1.
	\item \texttt{d2}: The true mean of difference of response for MRCT 2. Defaults to \texttt{d1}.  
	\item \texttt{sigmaTrt1}: The standard deviation of response in the treatment group for MRCT 1.
	\item \texttt{sigmaCtrl1}: The standard deviation of response in the control group for MRCT 1. Defaults to \texttt{sigmaTrt1}.
	\item \texttt{sigmaTrt2}: The standard deviation of response in the treatment group for MRCT 2. Defaults to \texttt{sigmaTrt1}.
	\item \texttt{sigmaCtrl2}: The standard deviation of response in the control group for MRCT 2. Defaults to \texttt{sigmaTrt2}.
	\item \texttt{randRatio1}: The randomization ratio between the treatment group and control group for MRCT 1. 
	\item \texttt{randRatio2}: The randomization ratio between the treatment group and control group for MRCT 2. Defaults to \texttt{randRatio1}.
\end{description}
\end{quote}

\begin{quote}
Detailed description of the arguments \texttt{conProb2} are as follows:
\begin{description}
	\item \texttt{alpha}: The Type I error.
	\item \texttt{power1}: Power for MRCT 1.
	\item \texttt{power2}: Power for MRCT 2. Defaults to \texttt{power1}.
	\item \texttt{pi}: The threshold ratio in the extended Japan’s criterion I (conditional version). Defaults to 0.5.
	\item \texttt{rF1}: The regional fraction for MRCT 1.
	\item \texttt{rF2}: The regional fraction for MRCT 2. Defaults to \texttt{rF1}.
	\item \texttt{d1}: The true mean of difference of response for MRCT 1.
	\item \texttt{d2}: The true mean of difference of response for MRCT 2. Defaults to \texttt{d1}.  
	\item \texttt{sigmaTrt1}: The standard deviation of response in the treatment group for MRCT 1.
	\item \texttt{sigmaCtrl1}: The standard deviation of response in the control group for MRCT 1. Defaults to \texttt{sigmaTrt1}.
	\item \texttt{sigmaTrt2}: The standard deviation of response in the treatment group for MRCT 2. Defaults to \texttt{sigmaTrt1}.
	\item \texttt{sigmaCtrl2}: The standard deviation of response in the control group for MRCT 2. Defaults to \texttt{sigmaTrt2}.
	\item \texttt{randRatio1}: The randomization ratio between the treatment group and control group for MRCT 1. 
	\item \texttt{randRatio2}: The randomization ratio between the treatment group and control group for MRCT 2. Defaults to \texttt{randRatio1}.
\end{description}
\end{quote}

A complete reference manual of the package is provided at \href{https://github.com/kunhaiq/ssMRCT.git}{https://github.com/kunhaiq/ssMRCT.git}.  

\end{document}